\documentclass[aps,prd,preprint,superscriptaddress,tightenlines,nofootinbib]{revtex4}
\usepackage{graphicx}
\usepackage{dcolumn}
\usepackage{bm}
\usepackage{epsfig}

\def\etal{et al.}

\newcommand{\Psidp}{$\psi(3770)$}
\newcommand{\Psip}{$\psi(2S)$}
\newcommand{\DDbar}{$D \bar{D}$}

\newcommand{\ee}{$e^{+} e^{-}$}

\newcommand{\KsKl}{$K_{S}^{0} K_{L}^{0}$}

\newcommand{\toKsKl}{$\to K_{S}^{0} K_{L}^{0}$}

\newcommand{\Ks}{$K_{S}^{0}$}
\newcommand{\KsKs}{$K_{S}^{0}K_{S}^{0}$}
\newcommand{\Kl}{$K_{L}^{0}$}
\newcommand{\pio}{$\pi^{0}$}

\newcommand{\KKstar}{$K^{{\ast} 0}(892) \bar{K^0}$}
\newcommand{\chic}{$\chi_c$}
\newcommand{\qed}{$e^{+} e^{-} \to \gamma \gamma,~l^{+} l^{-}$}

\begin{document}

\preprint{CLNS 05/1946}       
\preprint{CLEO 05-32}         

\title{Search for the non-\DDbar\ decay \Psidp\ \toKsKl}

\author{D.~Cronin-Hennessy}
\author{K.~Y.~Gao}
\author{D.~T.~Gong}
\author{J.~Hietala}
\author{Y.~Kubota}
\author{T.~Klein}
\author{B.~W.~Lang}
\author{R.~Poling}
\author{A.~W.~Scott}
\author{A.~Smith}
\affiliation{University of Minnesota, Minneapolis, Minnesota 55455}
\author{S.~Dobbs}
\author{Z.~Metreveli}
\author{K.~K.~Seth}
\author{A.~Tomaradze}
\author{P.~Zweber}
\affiliation{Northwestern University, Evanston, Illinois 60208}
\author{J.~Ernst}
\affiliation{State University of New York at Albany, Albany, New York 12222}
\author{H.~Severini}
\affiliation{University of Oklahoma, Norman, Oklahoma 73019}
\author{S.~A.~Dytman}
\author{W.~Love}
\author{S.~Mehrabyan}
\author{V.~Savinov}
\affiliation{University of Pittsburgh, Pittsburgh, Pennsylvania 15260}
\author{O.~Aquines}
\author{Z.~Li}
\author{A.~Lopez}
\author{H.~Mendez}
\author{J.~Ramirez}
\affiliation{University of Puerto Rico, Mayaguez, Puerto Rico 00681}
\author{G.~S.~Huang}
\author{D.~H.~Miller}
\author{V.~Pavlunin}
\author{B.~Sanghi}
\author{I.~P.~J.~Shipsey}
\author{B.~Xin}
\affiliation{Purdue University, West Lafayette, Indiana 47907}
\author{G.~S.~Adams}
\author{M.~Anderson}
\author{J.~P.~Cummings}
\author{I.~Danko}
\author{J.~Napolitano}
\affiliation{Rensselaer Polytechnic Institute, Troy, New York 12180}
\author{Q.~He}
\author{J.~Insler}
\author{H.~Muramatsu}
\author{C.~S.~Park}
\author{E.~H.~Thorndike}
\affiliation{University of Rochester, Rochester, New York 14627}
\author{T.~E.~Coan}
\author{Y.~S.~Gao}
\author{F.~Liu}
\author{R.~Stroynowski}
\affiliation{Southern Methodist University, Dallas, Texas 75275}
\author{M.~Artuso}
\author{S.~Blusk}
\author{J.~Butt}
\author{J.~Li}
\author{N.~Menaa}
\author{R.~Mountain}
\author{S.~Nisar}
\author{K.~Randrianarivony}
\author{R.~Redjimi}
\author{R.~Sia}
\author{T.~Skwarnicki}
\author{S.~Stone}
\author{J.~C.~Wang}
\author{K.~Zhang}
\affiliation{Syracuse University, Syracuse, New York 13244}
\author{S.~E.~Csorna}
\affiliation{Vanderbilt University, Nashville, Tennessee 37235}
\author{G.~Bonvicini}
\author{D.~Cinabro}
\author{M.~Dubrovin}
\author{A.~Lincoln}
\affiliation{Wayne State University, Detroit, Michigan 48202}
\author{D.~M.~Asner}
\author{K.~W.~Edwards}
\affiliation{Carleton University, Ottawa, Ontario, Canada K1S 5B6} 
\author{R.~A.~Briere}
\author{I.~Brock}~\altaffiliation{Current address: Universit\"at Bonn, Nussallee 12, D-53115 Bonn}
\author{J.~Chen}
\author{T.~Ferguson}
\author{G.~Tatishvili}
\author{H.~Vogel}
\author{M.~E.~Watkins}
\affiliation{Carnegie Mellon University, Pittsburgh, Pennsylvania 15213}
\author{J.~L.~Rosner}
\affiliation{Enrico Fermi Institute, University of
Chicago, Chicago, Illinois 60637}
\author{N.~E.~Adam}
\author{J.~P.~Alexander}
\author{K.~Berkelman}
\author{D.~G.~Cassel}
\author{J.~E.~Duboscq}
\author{K.~M.~Ecklund}
\author{R.~Ehrlich}
\author{L.~Fields}
\author{L.~Gibbons}
\author{R.~Gray}
\author{S.~W.~Gray}
\author{D.~L.~Hartill}
\author{B.~K.~Heltsley}
\author{D.~Hertz}
\author{C.~D.~Jones}
\author{J.~Kandaswamy}
\author{D.~L.~Kreinick}
\author{V.~E.~Kuznetsov}
\author{H.~Mahlke-Kr\"uger}
\author{T.~O.~Meyer}
\author{P.~U.~E.~Onyisi}
\author{J.~R.~Patterson}
\author{D.~Peterson}
\author{E.~A.~Phillips}
\author{J.~Pivarski}
\author{D.~Riley}
\author{A.~Ryd}
\author{A.~J.~Sadoff}
\author{H.~Schwarthoff}
\author{X.~Shi}
\author{S.~Stroiney}
\author{W.~M.~Sun}
\author{T.~Wilksen}
\author{M.~Weinberger}
\affiliation{Cornell University, Ithaca, New York 14853}
\author{S.~B.~Athar}
\author{P.~Avery}
\author{L.~Breva-Newell}
\author{R.~Patel}
\author{V.~Potlia}
\author{H.~Stoeck}
\author{J.~Yelton}
\affiliation{University of Florida, Gainesville, Florida 32611}
\author{P.~Rubin}
\affiliation{George Mason University, Fairfax, Virginia 22030}
\author{C.~Cawlfield}
\author{B.~I.~Eisenstein}
\author{I.~Karliner}
\author{D.~Kim}
\author{N.~Lowrey}
\author{P.~Naik}
\author{C.~Sedlack}
\author{M.~Selen}
\author{E.~J.~White}
\author{J.~Wiss}
\affiliation{University of Illinois, Urbana-Champaign, Illinois 61801}
\author{M.~R.~Shepherd}
\affiliation{Indiana University, Bloomington, Indiana 47405 }
\author{D.~Besson}
\affiliation{University of Kansas, Lawrence, Kansas 66045}
\author{T.~K.~Pedlar}
\affiliation{Luther College, Decorah, Iowa 52101}
\collaboration{CLEO Collaboration} 
\noaffiliation

\date{\today}

\begin{abstract} 
Using the current world's largest data sample of \Psidp\ decays, we present
results of a search for the non-\DDbar\ decay \Psidp\ \toKsKl. 
We find no signal, and
obtain an upper limit of $\sigma$(\Psidp\ \toKsKl)~$<~0.07 $~pb at
90\% confidence level (CL). Our result tests a theoretical
prediction for the upper bound on
$\cal{B}$(\Psidp\ \toKsKl) based on a charmonia-mixing model.
\end{abstract}

\pacs{13.20.Fc,13.20.Gd,12.38.Qk}
\maketitle

The \Psidp, the lightest charmonium resonance above open-flavor-production threshold, decays dominantly into a pair of $D $
mesons. Non-\DDbar\ decays of \Psidp, which are
OZI-suppressed~\cite{ozi}, have received much attention as they were
observed to constitute an unexpectedly large portion ($\approx$20\%) of the total hadronic decay rate in earlier
measurements summarized in~\cite{rosners latest2}.
Recent results~\cite{DDbar papers, hajime's paper} do not confirm a discrepancy of this magnitude, but due to the experimental uncertainties, the measurements are not yet conclusive. Exclusive non-\DDbar\ decays of the \Psidp\ are expected as of any
other charmonium state and have, in fact, been observed~\cite{brian
suggested,gamma chic1,gadams}. The discrepancy between the hadronic
and $D$-pair production cross sections will be clarified by
identifying other components of \Psidp\ $\to \mathrm{non}$-\DDbar.

This article focuses on the  reaction \Psidp\ \toKsKl. This
final state is particularly interesting also in the context of the
perturbative QCD ``12$\%$ rule'' (which relates $J/\psi$ and \Psip\
branching fractions~\cite{rule}), with the help of the mixing
scenario~\cite{modelsSD,mix1}, in which the \Psip\ and
\Psidp\ are considered to be mixtures of the $2^{3}S_{1}$ and $1^{3}D_{1}$ states of charmonium.
Rosner~\cite{ros_paper} propounds a possible explanation of
the ``$\rho-\pi $ puzzle'' using the phenomenon of charmonia mixing. 
In view of this, it is interesting to explore the
nature and degree of mixing between the \Psip\ and \Psidp,
particularly in the case of those modes which show significant
deviation from the 12\% rule. 
The pseudoscalar-pseudoscalar (PP) final state \KsKl\ is in this
category: an average of recent results~\cite{modelsSD,our paper}
indicates the ratio ${\cal B}$$(\psi(2S) \to K_{S}^{0} K_{L}^{0})/{\cal B}$$(J/\psi \to K_{S}^{0} K_{L}^{0})$, 
to be as high as $(29.9 \pm 3.0)$\%,
substantially enhanced compared to the prediction of 12$\%$.
In a scenario that attempts to explain this enhancement, Wang and
collaborators~\cite{mix1} predict 
$(1.2 \pm 0.7) \times 10^{-6} < \cal{B}$$(\psi(3770) \to
K_{S}^{0} K_{L}^{0}) <~(3.8 \pm 1.1) \times 10^{-5} $. The search
presented here has sufficient sensitivity to test the upper bound of this prediction.

We use data collected by the CLEO detector~\cite{cleo} operating 
at the Cornell Electron Storage Ring~\cite{cleoc}.
The sample corresponds to an integrated luminosity of
281~pb$^{-1} $ of \ee\ annihilations at a center-of-mass energy $\sqrt{s}= $3773~MeV.
The CLEO-c detector configuration features excellent efficiency and
resolution for charged particles and photons within 93\% of the solid
angle.
The detector components critical to this analysis, rendering the discrimination
of signal events amidst background, are discussed in the following.   
The charged particle tracking system consists of a low mass
wire inner drift chamber (ZD) suitable for low track momenta, followed by an outer drift chamber (DR).
These two devices measure charged track three-momenta with
excellent accuracy and achieve a momentum resolution of $\sim $0.6\%
at 1~GeV/c. The DR also measures energy loss that is used to
identify charged tracks.
The drift chamber is surrounded by a Ring Imaging Cherenkov
Detector (RICH), followed by a CsI calorimeter (CC), 
where two regions in polar angle (measured with respect to the beam
direction) are
distinguished: barrel ($|\cos(\theta)| < $ 0.81) and endcap ($|\cos(\theta)| \geq $ 0.81).
The CC allows the
detection of photons with an energy resolution of 2.2\% (5\%) for photons with
energy of 1~GeV (100~MeV) and, in combination with the tracking system,
provides the basis for excellent electron identification. 

The event reconstruction and final state selection criteria proceed along
the lines of CLEO's \Psip\ $\to $ \KsKl\ analysis~\cite{our
paper}. Our strategy is to reconstruct only a single \Ks\ in each
event and demand nothing other than an accompanying \Kl\ based on the following
criteria. 
We reconstruct the \Ks\ using its decay to
two charged pions, and thus require the events to have
exactly two charged tracks. We have taken into account the effect of the
small ($\approx$ 4 mrad) crossing angle between the $e^{+} $ and $e^{-} $ beams
by performing a Lorentz transformation of all the laboratory quantities to the
center-of-mass frame.  
We impose standard track selection criteria based on the number of
drift chamber hits and geometric acceptance.
We use both charged particle ionization loss in the drift
chamber ($dE/dx $) and RICH information to identify the two tracks as
pions which are used to reconstruct the \Ks\ mesons. We define the
parameters ${\mathrm {PID}}_{1 i j}= L_i - L_j$~and ${\mathrm {PID}}_{2 i j} =
\sigma^2_i - \sigma^2_j$, where $L_{i,j}(i,j = p, K, \pi~{\rm with}~i \neq j)$~are the
likelihoods given by the measured Cherenkov angles of photons in the RICH detector compared
with the predicted Cherenkov angles for the $i^{\rm th},j^{\rm th}$~particle hypothesis,
and $\sigma_{i,j}(i,j = p, K, \pi, e~{\rm with}~i \neq j)$~are the ratios of the difference between the measured
$dE/dx$~and the predicted $dE/dx$~values normalized to their standard
deviations for the $i^{\rm th}, j^{\rm th}$~hypothesis.
To use the RICH, we first require that the
track momentum be above the RICH threshold of ${\rm 0.6~GeV/c}$~for the
pion hypothesis. We then discriminate $\pi $ from $p, K $ in the following manner: (a) If the RICH
information is available, we require 
$({\mathrm {PID}}_{1 \pi p} + {\mathrm {PID}}_{2 \pi p}) <~0$ and 
$({\mathrm {PID}}_{1 \pi K} + {\mathrm {PID}}_{2 \pi K}) <~0$. (b) If the RICH information is not available, we require ${\mathrm {PID}}_{2 \pi p} <~0$~and
${\mathrm {PID}}_{2 \pi K} <~0$.
We further reject background from Bhabha (\ee\ $\to $ \ee) and two-photon
(\ee\ $\to \gamma\gamma$) events by an
electron veto of tracks that satisfy the condition for the ratio of
the CC determined energy $E_{{\rm CC}}$ and the track determined
momentum $p$ as $0.92 <E_{{\rm CC}}/p<~1.05$~and have $|\sigma_{e}| <~3$.

The pair of charged pion candidates are kinematically constrained
to come from a common vertex. We
require that the reconstructed invariant mass of the two pions be within 10~MeV
($\approx 3.2$ standard deviations) of the nominal \Ks\ peak. We reject background from 
non-\Ks\ sources by requiring the measured flight path of the \Ks\ candidate 
before its decay to be greater than 5 standard deviations 
($\approx 5$~mm) with respect to the interaction point. In addition, 
we require that the \Ks\ candidates originate from the $e^+e^-$
interaction point by demanding that their distance of closest
approach is within 5 standard deviations with respect to the interaction
point. 
 
Frequently, \Kl\ mesons will produce a shower in the CC. However, as
this shower does not have a measured energy corresponding to the energy of the \Kl, we do not attempt to reconstruct it, but merely allow for its existence.
In order to reject contamination from anticipated background events
with energy from neutral particles other than the \Kl, 
we impose selection conditions as follows.
We first find the \Kl\ direction
by looking opposite to the \Ks\ direction, after having
constrained the \Ks\ to its nominal mass and also considered the effect
of the finite crossing angle.
We then require the energy of the shower associated with neutrals 
closest to the \Kl\ direction to be less than 1.5~GeV. This
cut further rejects QED background
events of the type $e^{+} e^{-} \to \gamma \gamma$ (where one of the
$\gamma $ undergoes pair production). 
We also require the sum of the energy associated with neutrals 
outside a region around the direction of the \Kl, defined by the 
angle between the position vectors of the \Kl\ and the shower in
consideration ($\delta_{\mathrm{angle}}> $ 0.35~radians), to be less than 300~MeV. We do not include the energy from showers which have an energy below
50~MeV, which are frequently due to electronic noise, in this 
summation.
This cut is very efficient in eliminating hadronic background
events from the following sources: 
\begin{enumerate}
\item \KKstar\ $+$ {\it {\rm charge conjugate}} ($\mathrm{c.c.}$)~produced in:
\begin{enumerate}
\item \ee\ $\to~\gamma$~\Psip, followed by the subsequent decay 
\Psip\ $\to $ \KKstar\ $+~\mathrm{c.c.}$, 
\item \ee $\to $ \KKstar\ $+~\mathrm{c.c.}$~via $e^+e^-\to\gamma^*$
(continuum). 
\end{enumerate}
\item $K_{S}^{0} K_{S}^{0}$ produced in: 
\begin{enumerate}
\item \ee $\to\gamma$~\Psip, followed by the subsequent radiative decay 
\Psip\ $\to \gamma$\chic$_{0,2}$, \chic$_{0,2}$ $\to K_{S}^{0}K_{S}^{0}$, 
\item \ee $\to$
\Psidp\ $\to \gamma $ \chic$_{0,2} $, \chic$_{0,2}$ $\to K_{S}^{0}K_{S}^{0}$,
\item \ee\ $\to K_{S}^{0} K_{S}^{0} $ via $e^+e^- \to \gamma^*$ (continuum).
\end{enumerate} 
\end{enumerate}
We define a ``good'' shower as one that has an energy profile 
consistent with being a photon and possessing an energy above 100~MeV,
and require events to have no good showers
associated with neutrals outside the \Kl\ region, and at most one
good shower inside this region.

Most events that contain one or more $\pi^0$ decays will be
eliminated by the above cuts. However, for even better rejection, we find it 
useful to introduce an explicit \pio\ veto with a lower photon energy requirement.
To identify a \pio, we require a pair of showers not associated with 
charged tracks to have their 
energy distribution consistent with a photon even if they overlap with
nearby clusters. In addition, we require each photon shower candidate
to possess at least 30~MeV (50~MeV) of energy for a barrel (endcap) 
photon and kinematically constrain the pair to the known \pio\ mass. We
further require the difference between the
unconstrained and fitted \pio\ mass, normalized by its resolution, to be $\leq
$ 3 and the \pio\ momentum to be $>$ 100~MeV/c. 
We reject all events that have any \pio\ candidates meeting the above criteria.
This is based
upon our understanding of the basic topology of the event and our
anticipated background interactions which produce \pio\ mesons. 
Many background events will fail more than one
of these cuts, but for best background rejection, all are necessary.

The above selection conditions were optimized by studying a Monte
Carlo simulation of events using the {\sc evtgen} generator~\cite{evtgen}
and a {\sc geant}-based~\cite{geant} detector modeling program. We simulated events
with a $\sin^2\theta$ angular distribution, where 
$\theta$ is the angle between the \Ks\ and the positron beam in the 
center-of-mass system, as is expected for a vector resonance decaying
to two pseudoscalar mesons. We also included initial state radiation effects.

\begin{figure}
\includegraphics*[width=3.5in]{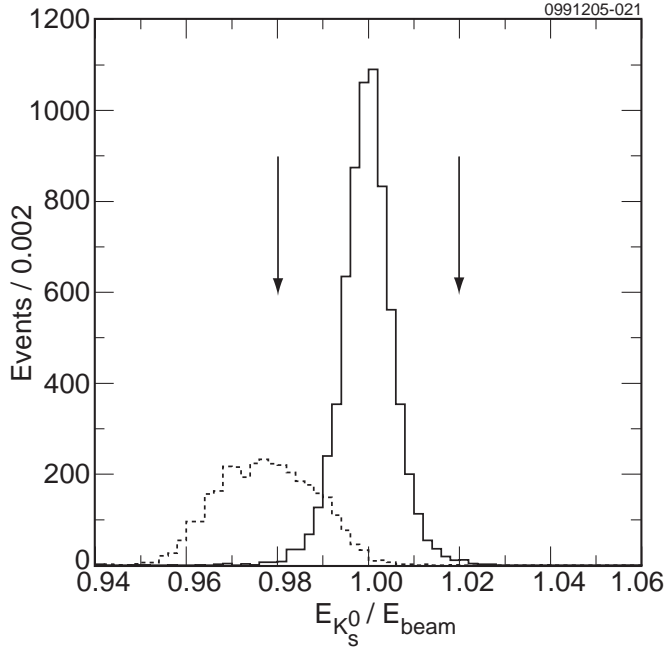}
\caption{The scaled $K_{S}^{0} $ energy
($E_{K_{S}^{0}}/E_{\mathrm{beam}}$) distribution for the $\psi(3770)$
simulation samples corresponding to the signal (solid histogram) and radiative
return (dotted histogram) processes using arbitrary normalization. 
The arrows mark the signal region.} 
\label{fig:MC}
\end{figure}

We define 
a {\rm scaled energy} variable for each
event as the ratio of the \Ks\ energy to the beam energy as
$E_{K_{S}^{0}}/E_{\mathrm{beam}} $. Since the \Ks\ mesons from the
signal will be mono-energetic, we
expect the $E_{K_{S}^{0}}/E_{\mathrm{beam}} $ distribution to peak at unity for the
signal. Based upon
simulations (as seen in Figure~\ref{fig:MC}), we determine the signal
region to be $0.98 < E_{K_{S}^{0}}/E_{\mathrm{beam}} < 1.02 $.

The relevant backgrounds studied are the QED sources (\qed) and the
hadronic sources: (a) \KKstar\ $+~\mathrm{c.c.}$~where the
$K^{{\ast}0}(892) $ decays into \pio\ and $K^0$, giving two neutral
kaons in the final state, which could become a \Ks\ and a \Kl, and 
(b) \KsKs\ where one of the \Ks\ $\to \pi^{+} \pi^{-} $,
while the other \Ks\ $\to \pi^{0} \pi^{0} $ and thus can mimic the signal. 
Two additional sources of background are
identified and studied. One of them is $D \bar D$ production.
Our Monte Carlo simulation study of a generic \DDbar\ sample more
than twice as large as the data indicates that this background is completely eliminated by our selection criteria.   
The second source of hadronic backgrounds taken into account is the
more pervasive \KsKl\ events originating from \Psip~\cite{our paper} 
produced at 3773~MeV in either the tail of
or the radiative return to the \Psip, the latter of which
is the bigger contribution.
These events peak at $E_{K_{S}^{0}}/E_{\mathrm{beam}}=0.977$ 
as seen in Figure~\ref{fig:MC}, which uses a
simulated sample of such events analyzed in an identical manner to the data. 
Unfortunately, these events cannot be eliminated fully in this analysis by
using the total four-momentum constraint typically used in \Psidp\
studies, as in this analysis we do not reconstruct the complete event.

\begin{figure}
\includegraphics*[width=3.5in]{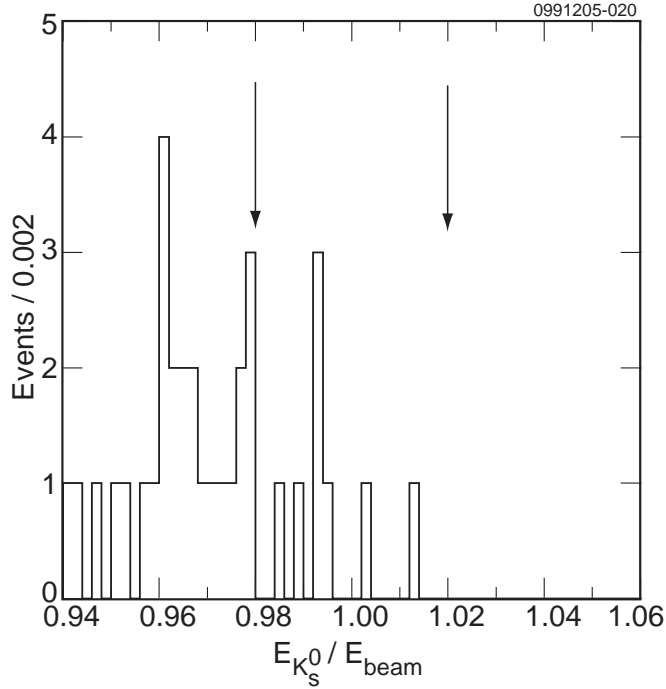}
\caption{The scaled $K_{S}^{0} $ energy ($E_{K_{S}^{0}}/E_{\mathrm{beam}}$)
distribution for the $\psi(3770) \rightarrow$ \KsKl\ decay in 281~pb$^{-1} $ of data. The arrows mark the signal region.}
\label{fig:Data}
\end{figure}

Figure~\ref{fig:Data} shows the $E_{K_{S}^{0}}/E_{\mathrm{beam}} $ distribution for
events in data after the application of all of the above selection cuts.
We observe 8 events inside the signal region, and an asymmetric
background in the low sideband $E_{K_{S}}/E_{\mathrm{beam}} < $ 0.98. Simulation studies indicate that the events in the low
sideband are not from backgrounds to the \Ks\ sidebands but rather
from multiple hadronic sources, dominated by contamination from
$\gamma$~\Psip $\to \gamma$~\KsKl, 
which may well extend inside the signal region.
In order to account for this, we estimate the number of events
expected inside our signal region originating from this radiative
return background.
Our estimate of the amount of this contamination, obtained using the 
radiative return hadronic cross section at $\sqrt s =$~3773~MeV~\cite{brian suggested} and 
$\cal{B}$(\Psip\ \toKsKl)~\cite{our paper}, is 9.5$\pm $1.6 events inside the
signal region. The uncertainty on the background estimate is found
to be 17.1\% by taking the quadrature sum of the relevant
sources which include the uncertainties on the radiative return cross
section~\cite{brian suggested}, the \Psidp\ luminosity, the detection
efficiency as obtained from simulations (the first six components
listed in Table~\ref{tab:sys}), $\cal{B}$(\Psip\ $\to$~\Ks\Kl)~\cite{our paper} and $\cal{B}$(\Ks\ $\to \pi^+ \pi^-$)~\cite{PDG 2004}. 
From such estimates, we find that the radiative return is a major part of
the background, and it saturates our data in the signal region.
In this analysis, we do not consider continuum subtraction from the
resonance yield as this final state is 
forbidden to be produced via electromagnetic interactions under the SU(3)
symmetry of flavor~\cite{Hab perier}, as confirmed in~\cite{our paper}.  

The detection efficiency for the signal channel is calculated to be
45.8\% from Monte Carlo simulations.
This efficiency includes a correction due to the difference in \Kl\
detection efficiencies determined between data and Monte Carlo simulation.
The study uses \ee\ $\to \gamma \phi$, $\phi \to $ \KsKl\ events
from the continuum channel in
the same data sample as used in the analysis of \Psidp\ $\to $ \KsKl.
A possible momentum dependence arising from a difference in the
momentum spectra of the \Kl\ mesons from the continuum and resonance
channels is accounted for in the form of a systematic error component
in the overall error attributed to the \Kl\ selection of 3.7\%.
A hardware trigger requiring that two tracks be found within 20 cm of
the event vertex eliminated very long-lived \Ks\ mesons. The
efficiency of this trigger for signal events was 73.6\%, which is
included in the calculated detection efficiency.

Based on our observed signal yield and estimated radiative return
background, we find no evidence for a signal and calculate an upper limit on the number of
signal events using the Feldman-Cousins
approach~\cite{Feldman-Cousins} for an observed yield of 8 and a
background estimate of 9 events which translates into an upper limit
(90\% CL) on the cross section of $\sigma (\psi(3770) \to
K_{S}^{0} K_{L}^{0} ) <$~0.06~pb.
In order to include systematic uncertainties,
we first consider the systematic uncertainty on the estimated background. We take the 1$\sigma $ lower value of
this estimate and calculate the 90\% CL upper limit on the number of
signal events using the Feldman-Cousins approach.
We subsequently include an overall systematic uncertainty of 5.5\%
from various sources listed in Table~\ref{tab:sys}, and measure
$\sigma (\psi(3770) \to K_{S}^{0} K_{L}^{0} ) < $ 0.07~pb at 90\% CL.

 \begin{table}[ht]
 \begin{center}
 \begin{tabular}{|c|c|}
 
 \hline
 \hline
 Source                                       &Systematic error (\%) \\
 \hline
                                              &                      \\
 Trigger                                      &2.0                      \\
 Simulation statistics                                &0.6\\
 \Kl\ selection				      &3.7\\	
 Tracking	                              &1.4                      \\
 Particle ID                                  &0.6\\
 \Ks\ finding	                              &3.0\\
 Luminosity                                   &1.0\\
 $\cal{B}$(\Ks\ $\to \pi^{+}\pi^{-} $)~\cite{PDG 2004}   &0.1\\
 				              &\\
 \hline					
 Total                                        &5.5\\
	
 \hline	
 \hline
 \end{tabular}
 \caption{Summary of systematic errors.}
 \label{tab:sys}
 \end{center}
 \end{table}

In conclusion,
using data collected by the CLEO detector at the
\Psidp\ resonance, we have studied the exclusive non-\DDbar\ 
\Psidp\ decay to the PP two-body mode \KsKl. We find no evidence of a signal
in 281~pb$^{-1} $ of data and report an upper limit on the production
cross section, $\sigma (\psi(3770) \to K_{S}^{0} K_{L}^{0} ) < 0.07 $~pb  at 90\% CL.
This indicates that this mode is far from being able to account for
any possible significant non-\DDbar\ rate from the 
\Psidp~\cite{rosners latest2,hajime's paper}.
In order to compare this result with the model prediction~\cite{mix1}, we
translate our upper limit on the cross section into an upper limit on
the branching fraction using the resonance cross section of the \Psidp\
reported by CLEO~\cite{hajime's paper}. We obtain {$\cal{B}$}($\psi(3770)
\to K_{S}^{0}K_{L}^{0}$)~$ < 1.17 \times 10^{-5} $ at 90\% CL, which
improves upon the current upper limit~\cite{BES psi3770 paper} of $<
2.1 \times 10^{-4} $ (90\% CL) by an order of magnitude.
Our measurement is below the upper bound of the model
prediction presented in ~\cite{mix1}. This may provide new insight
into the nature of $S$- and $D$-wave mixing in charmonia and thereby
help clarify the perturbative 12\% rule.

We gratefully acknowledge the effort of the CESR staff 
in providing us with excellent luminosity and running conditions.
This work was supported by 
the A.P.~Sloan Foundation,
the National Science Foundation,
and the U.S. Department of Energy.

\end{document}